\documentclass[12pt,aps,preprint,nofootinbib]{revtex4}

\usepackage{graphicx}
\usepackage{cancel}
\usepackage{amssymb}
\usepackage{textcomp}
\usepackage{amsmath}
\usepackage{bm}
\usepackage{times}
\usepackage{epsfig}
\usepackage{color}

\usepackage{comment}

\def\lsim{\mathrel{\raise.3ex\hbox{$<$\kern-.75em\lower1ex\hbox{$\sim$}}}}
\def\gsim{\mathrel{\raise.3ex\hbox{$>$\kern-.75em\lower1ex\hbox{$\sim$}}}}

\def\beq{\begin{equation}}
\def\eeq{\end{equation}}
\def\be{\begin{equation}}
\def\ee{\end{equation}}
\def\bea{\begin{eqnarray}}
\def\eea{\end{eqnarray}}

\def\etmiss{\cancel{E}_{T}}

\def\to{\rightarrow}

\newcommand{\minigraph}[5][0.25in]{\begin{minipage}{#2}\begin{center}\includegraphics[width=#2]{#5}\\\vspace{#3}\hspace{#1}{\footnotesize #4}\end{center}\end{minipage}}

\begin{document}
 
\title{
 Electroweakino Searches: A Comparative Study for LHC and ILC\\
(A Snowmass White Paper)
}

\author{Mikael Berggren$^{\bf a}$}
\email{mikael.berggren@cern.ch}

\author{Tao Han$^{\bf b}$}
\email{than@pitt.edu}

\author{Jenny List$^{\bf a}$} 
\email{jenny.list@desy.de} 

\author{Sanjay Padhi$^{\bf c}$}
\email{Sanjay.Padhi@cern.ch}

\author{Shufang Su$^{\bf d}$}
\email{shufang@email.arizona.edu}

\author{Tomohiko Tanabe$^{\bf e}$}
\email{tomohiko@icepp.s.u-tokyo.ac.jp}

\affiliation{
$^{\bf a}$  DESY, Notkestra{\ss}e 85, 22607 Hamburg, Germany \\
$^{\bf b}$  Pittsburgh Particle Physics, Astrophysics, and Cosmology Center, Department of Physics $\&$ Astronomy, University of Pittsburgh, 3941 O'Hara St., Pittsburgh, PA 15260, USA\\
$^{\bf c}$  Department of Physics, University of California--San Diego, 9500 Gilman Dr., La Jolla, CA 92093, USA\\
$^{\bf d}$  Department of Physics, University of Arizona, 1118 E. 4th St., Tucson, AZ 85721, USA \\
$^{\bf e}$  ICEPP, The University of Tokyo, Bunkyo-ku, Tokyo 113-0033, Japan
}

\begin{abstract}

We make a systematic and comparative study for the LHC and ILC for the electroweakino searches in the Minimal Supersymmetric Standard Model. We adopt a general bottom-up approach and scan over the parameter regions for all the three cases of the lightest supersymmetric particle being Bino-, Wino-, and Higgsino-like.    The electroweakino signal from pair production and subsequent decay to $Wh\ (h\to b\bar b)$ final state may yield a sensitivity of $95\%$ C.L.~exclusion (5$\sigma$ discovery) to the mass scale $M_{2},\ \mu \sim 250-400$ GeV ($200-250$ GeV) at the 14 TeV LHC with an luminosity of 300 fb$^{-1}$.
Combining with all the other decay channels, the $95\%$ C.L.~exclusion (5$\sigma$ discovery) may be extended to $M_{2},\ \mu \sim 480-700$ GeV ($320-500$ GeV). At the ILC, the electroweakinos could be readily discovered once the kinematical threshold is crossed, and their properties could be thoroughly studied. 

\end{abstract}

\maketitle

\section{Introduction}

The discovery of a Standard Model (SM)-like Higgs boson~\cite{ATLASH,CMSH} has further strengthened the belief for a weakly-coupled Higgs sector with supersymmetry (SUSY)
 as the most compelling realization. 
 If the weak-scale SUSY is realized in nature~\cite{Nilles:1983ge}, the definitive confirmation will require the discovery of the supersymmetric partners of the electroweak (EW) particles in the SM,  in particular the gauginos and Higgsinos. We refer to them as electroweakinos (EWkinos) throughout this document. 
 
Given the current results on SUSY searches at the LHC  \cite{ATLAS2013047,CMS2013012, ATLAS2013028, ATLAS2013035,ATLAS2013036, ATLAS2013049,ATLAS2013093,CMS2013006,CMS2013017}, the absence of the spectacular events of large hadronic activities plus substantial missing energy implies
 a lower mass bound for the gluino and light squarks at about a TeV or heavier.
 In anticipation of much heavier colored SUSY partners,  we are thus led to consider a more challenging search strategy, namely the SUSY signals only from the EW sector, the charginos and neutralinos.

Unfortunately, the direct production of electroweakinos
 at the LHC suffers  from relatively small rates  \cite{Baer:1994nr}. The current direct search bounds at the LHC are thus rather weak~\cite{ATLAS2013028, ATLAS2013035,ATLAS2013036, ATLAS2013049,ATLAS2013093,CMS2013006, CMS2013017} and the future perspectives for the mass parameter coverage are limited~ \cite{ATLASTDR, CMSTDR}. 
A further complication is that some dark matter consideration favors a situation for nearly-degenerate charginos and  neutralinos~\cite{Arkani-Hamed:2006mb}, making their identification more challenging~\cite{Giudice:2010wb}.
On the other hand, a lepton collider will provide a rather clean environment for new physics discovery, as long as the collider achieves higher enough luminosity and the center-of-mass (CM) energy crosses the new physics threshold. 

We are thus motivated to assess the sensitivity for observing the electroweakinos at the LHC and to compare with the reach at an ILC with a CM energy of 500 GeV.
The most relevant soft SUSY-breaking mass parameters for the Bino, Wino, and Higgsino are $M_1, M_{2}$ and $\mu$, respectively.
We take a model-independent approach and study the SUSY signals with all possible relative values of these three SUSY-breaking mass parameters, which lead to six cases in the most general terms \cite{THSPSS}.
 Among them, four cases would naturally result in a compressed spectrum of nearly degenerate lightest supersymmetric particles (LSPs).
Beyond the existing literature for studying the production and decay of the charginos and neutralinos, 
we also put a special emphasis on the by-now established channels from the decays of a 125 GeV Higgs boson $h\to b\bar b, WW^{*}$. 
We find that  with 300 fb$^{-1}$ of integrated luminosity at the 14 TeV LHC, it is promising to  reach up to an electroweakino mass about 400 GeV (250 GeV) for a $95\%$ C.L. exclusion (5$\sigma$ discovery), 
and the signal with $h\to b \bar b$ can be identified. At the ILC, once the mass threshold is reached for the gaugino pair production, the signal of charginos and neutralinos may be separated with $O(1\%)$ measurements for the mass resolution and the couplings. 
 

\section{Model Parameters and Neutralino/chargino decays}
\label{sec:model}

 We focus on an essential EW sector, namely, the gauginos and Higginos as the interaction eigenstates. 
Without assumptions for a SUSY-breaking mediation scenario,   we consider the other SUSY particles, namely, gluinos, squarks and sleptons,   as well as heavy Higgs bosons, to decouple from the spectrum.
We explicitly incorporate a SM-like Higgs boson of mass:
$m_{h} = 125\ {\rm GeV}$.  There are only four parameters involved in the electroweakino mass matrices, two soft SUSY breaking mass parameters $M_1$ and $M_2$, the Higgs field mixing parameter $\mu$, and the electroweak symmetry breaking parameter $\tan\beta$. 
Diagonalization of the mass matrices gives the mass eigenstates (with increasing mass eigenvalues), namely, the Majorana fermions neutralinos 
$\chi_i^0$ ($i=1 \ldots 4$),  and the Dirac fermions charginos $\chi_i^\pm$ ($i=1,2$).
  
For our phenomenological considerations,
 we adopt the parameters in the broad range
\begin{equation}
 100\ {\rm GeV}  < M_{1},\ M_2,\ |\mu| < 1~{\rm TeV}, \quad 3 < \tan{\beta} < 50.
\label{mixi}
\end{equation}
Note that in the case of Bino-like LSP, $M_1$ could be much lower given the lack of model-independent limit on the Bino mass.
 

\begin{figure}
\includegraphics[scale=1.5,width=5 in]{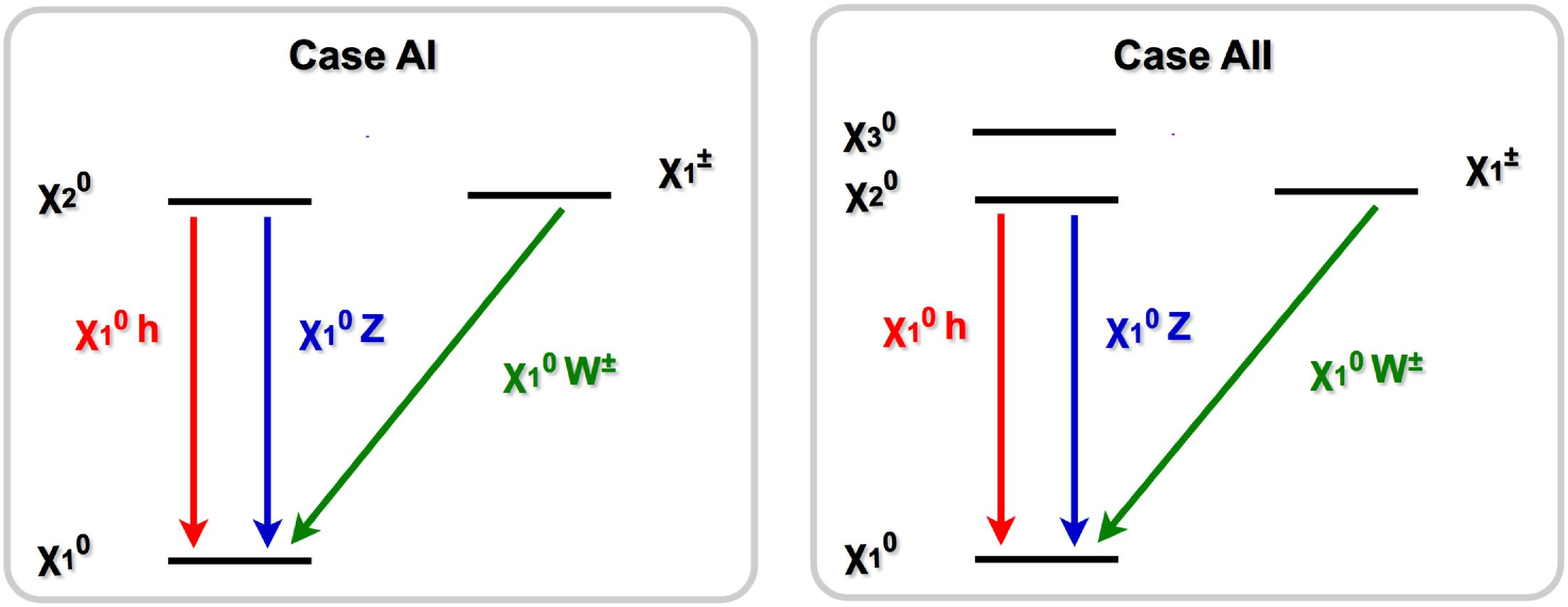}
\includegraphics[scale=1.5,width=5 in]{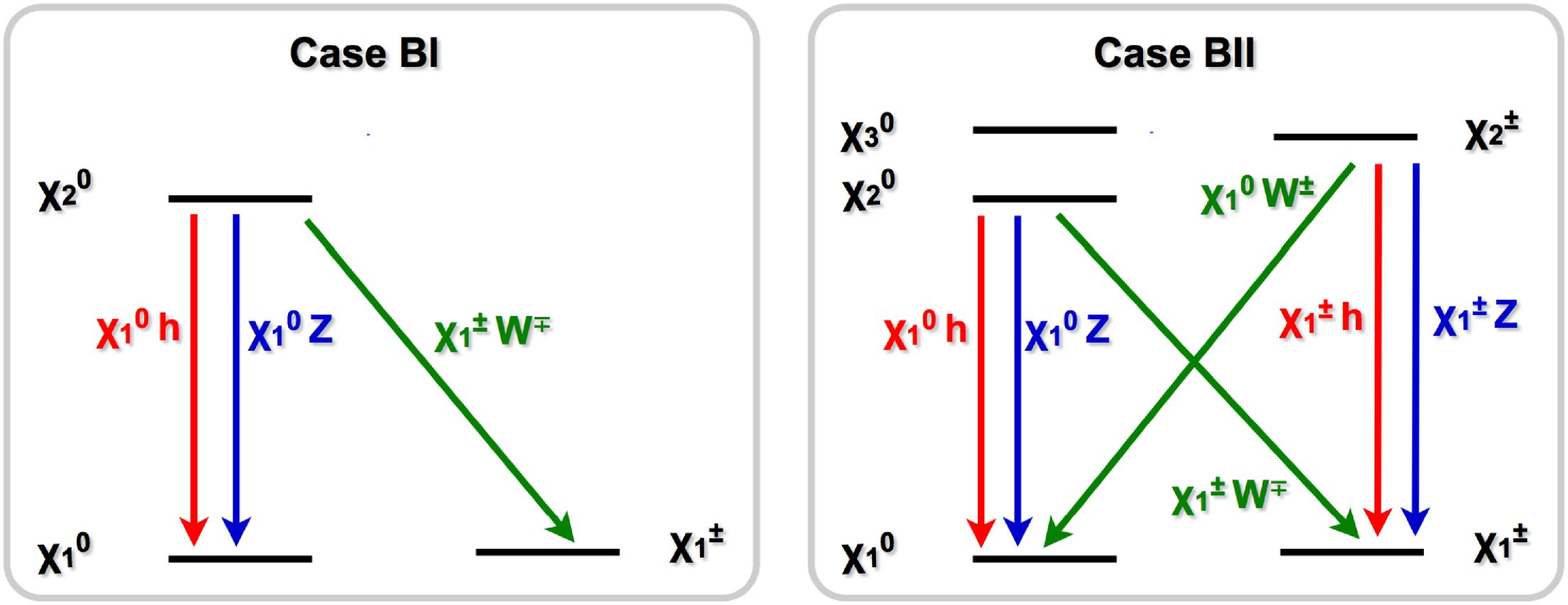}
\includegraphics[scale=1.5,width=5 in]{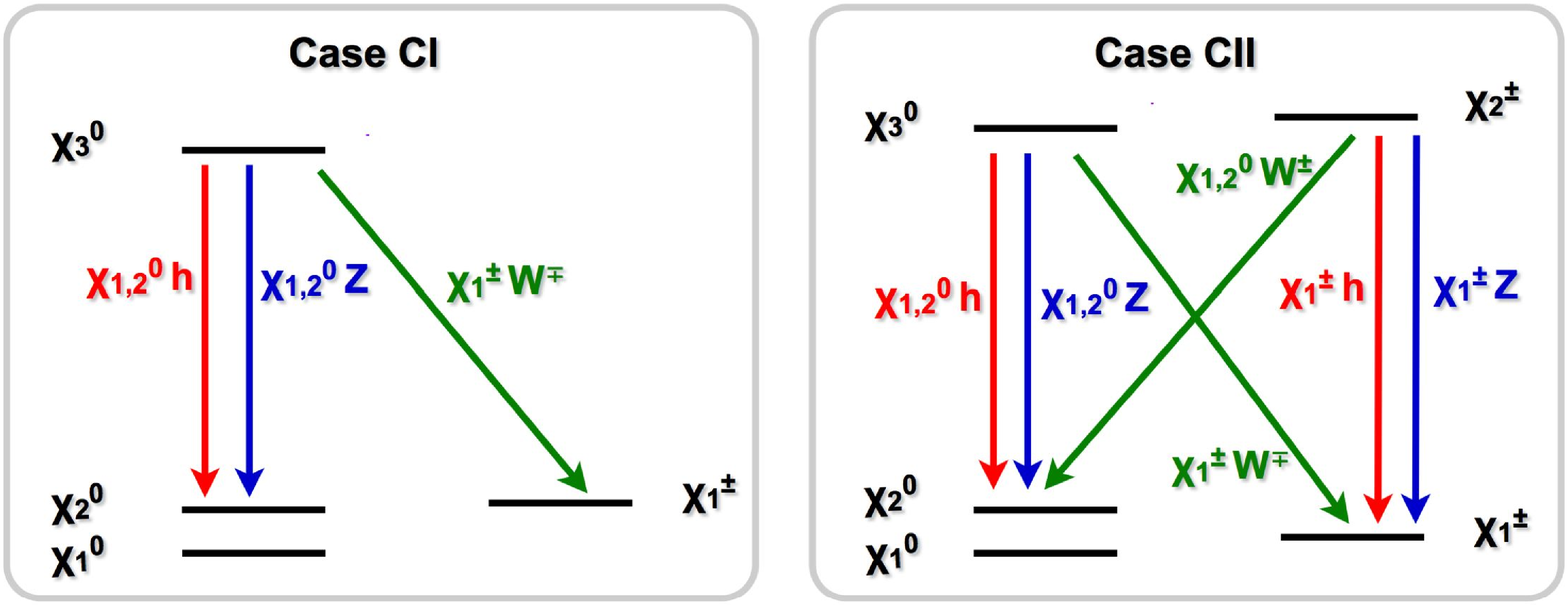}
\caption{Decay patterns of NLSP's for all the six cases AI$-$CII. } 
\label{fig:CaseABC_decay}
\end{figure}

To explore the phenomenological consequences in a most general approach, we present the three possible scenarios among the mass parameters of $M_{1},\ M_2,\ \mu$, and categorize them into six different cases. Each of those leads to characteristic phenomenology in their pair production and the decays of the charginos and neutralinos  \cite{THSPSS}.
 \begin{itemize}
{\bf \item{Scenario A:}  $M_{1} < M_{2},\ |\mu|$ }
\end{itemize}

This is the usual canonical scenario, which is strongly motivated by the Bino-like LSP dark matter 
 and by the grand unified theories 
with gaugino mass unification \cite{SUSYGUTs}. 
 There are two qualitatively different physics cases we would like to explore, namely
\bea
&& {\rm Case\ AI:}\quad M_{2} < |\mu|,\quad  \chi_1^\pm, \chi_2^0 {\rm \  \ are\ Wino-like};\  \chi_2^\pm, \chi_{3,4}^0 {\rm \ \ are\ Higgino-like}; \\
&& {\rm Case\ AII:}\quad |\mu| < M_{2},\quad  \chi_1^\pm, \chi_{2,3}^0 {\rm \ \ are\ Higgino-like};\ \chi_2^\pm, \chi_4^0 {\rm \  \ are\ Wino-like}.
\eea
 

\begin{itemize}
\item{\bf Scenario B:} $M_{2} < M_{1},\ |\mu|$
\end{itemize}

This is the situation of Wino-like LSP, as often realized in anomaly mediated SUSY breaking scenarios \cite{amsb}.
The lightest states $\chi_1^0$ and $\chi_1^\pm$ are nearly degenerate in mass close to $M_{2}$. It thus makes more sense to follow the newly introduced convention to call them all  ``LSPs''.
In this scenario, there are two possible mass relations we will explore
\bea
&& {\rm Case\ BI:}\quad M_{1} < |\mu|,\quad 
\ \chi_2^0 {\rm \  \ Bino-like}; \ \chi_2^\pm,\ \chi_{3,4}^0 {\rm \  \ Higgsino-like}; \\
&& {\rm Case\ BII:}\quad |\mu| < M_{1}, \quad  
 \ \chi_2^\pm,\ \chi_{2,3}^0 {\rm \  \ Higgsino-like}; \
 \chi_{4}^0 {\rm \ \  Bino-like}.~~
\eea
 

\begin{itemize}
\item{\bf Scenario C:} $|\mu| < M_{1},\ M_{2}$
\end{itemize}

This is the situation of Higgsino-like LSP
with the lightest states $\chi_{1,2}^0$ and $\chi_1^\pm$ being Higgsino-like.  The two possible mass relations here are 
\bea
&& {\rm Case\ CI:}\quad M_{1} < M_{2},\quad 
\  
\chi_{3}^0 {\rm \ \  Bino-like};\ \chi_2^\pm, \ \chi_4^0{\rm \  \ Wino-like};  
\\
&& {\rm Case\ CII:}\quad M_{2} < M_{1}, \quad  
 \chi_2^\pm,\ \chi_3^0{\rm \  \ Wino-like}; \  \chi_{4}^0 {\rm \ \  Bino-like}.~~
\eea
 

\begin{table}
 \begin{tabular}{|c| ll| c| c|c|c|c|c|c |c|}    \hline
    
   & \multicolumn{2}{|c|}{NLSP decay Br's }  & Production &\multicolumn{7}{|c|}{Total Branching Fractions (\%)}\\ \cline{5-11}
  
  &&&&$W^+W^-$&$W^\pm W^\pm$&$WZ$&$Wh$&$Zh$&$ZZ$&$hh$ \\
    \hline \hline
 Case AI & $\chi_1^\pm \rightarrow  \chi_1^0 W^\pm$ &100\%& $ \chi_1^\pm \chi_2^0 $  & 
 &&18&82&&&\\
$M_{1} < M_{2}< \mu$ & $\chi_2^0 \rightarrow \chi_1^0 h$ & 82\%(96$-$70\%) & $ \chi_1^+ \chi_1^-  $ & 
100 &&&&&&\\
\hline 
 Case AII  &$\chi_1^\pm \rightarrow  \chi_1^0 W^\pm$&100\%& $\chi_1^\pm \chi_{2}^0$&
 &&26&74&&&\\
$M_{1} <\mu< M_{2}$  & $\chi_2^0 \rightarrow \chi_1^0 h$  &74\%(90$-$70\%)& $\chi_1^\pm \chi_{3}^0$ &
 &&78&23&&&\\
  
 & $\chi_3^0 \rightarrow \chi_1^0 Z$&78\%(90$-$70\%)&$\chi_1^+ \chi_1^-$&
 100 &&&&&&\\
 &  &  &
$ \chi_2^0\chi_3^0 $&
 &&&&63&20&17\\
 
  \hline \hline

Case BI &\multicolumn{10}{|l|}{ }\\ 
 $M_{2} < M_{1}< \mu$ &\multicolumn{10}{|l|}{$\chi_2^0\rightarrow \chi_1^\pm W^\mp, \chi_1^0 h, \chi_1^0 Z$, \ \ \ 68\%, 27\%($31-24\%$), 5\%($1-9\%$), \ \ \ production suppressed.}\\ \hline
 Case BII &$\chi_2^\pm \rightarrow  \chi_1^0 W^\pm$ &35\%   &$\chi_2^\pm \chi_{2}^0$&
12&12&32&23&10&9&2
\\   
 $M_{2} < \mu < M_{1}$ &$\chi_2^\pm \rightarrow \chi_1^\pm Z$&35\%   &$\chi_2^\pm \chi_{3}^0$ &
 12&12&26&29&11&3&7
  \\   
  &$\chi_2^\pm \rightarrow \chi_1^\pm h$ & 30\%   &$\chi_2^+ \chi_{2}^-$&
 12&&25&21&21&12&9
  \\  
    
  &$\chi_2^0 \rightarrow \chi_1^\pm W^\mp$ &67\%   &$\chi_2^0 \chi_{3}^0$&
23&23&23&21&7&2&2
 \\   
  &$\chi_2^0 \rightarrow \chi_1^0 Z$ & 26\%(30$-$24\%)  &&
&&&&&&
  \\
 &$\chi_3^0 \rightarrow \chi_1^\pm W^\mp$ &68\%   &&  
 &&&&&&\\
&$\chi_3^0 \rightarrow \chi_1^0 h$ &24\%(30$-$23\%) &&
 &&&&&&\\
 \hline \hline
Case CI &\multicolumn{10}{|l|}{ }\\ 
$\mu < M_{1} < M_{2}$  &\multicolumn{10}{|l|}{$\chi_3^0\rightarrow \chi_1^\pm W^\mp, \chi_{1,2}^0 Z, \chi_{1,2}^0 h$, \ \ \ 52\%, 26\%, 22\%, \ \ \ production suppressed.}\\ \hline
 Case CII &$\chi_{2}^\pm \rightarrow  \chi_{1,2}^0 W^\pm$ &51 \%   &$\chi_2^\pm \chi_{3}^0$&
14&14&27&23&11&6&5
\\   
$\mu < M_{2} < M_{1}$  &$\chi_{2}^\pm \rightarrow  \chi_{1}^\pm Z$ &26 \%   &$\chi_2^+ \chi_{2}^-$&
26&&26&24&12&7&5
\\   
  &$\chi_{2}^\pm \rightarrow  \chi_{1}^\pm h$ &23 \%   & &
&&&&&&
\\   
  &$\chi_{3}^0 \rightarrow  \chi_{1}^\pm W^\mp$ &54 \%   & &
&&&&&&
\\   
  &$\chi_{3}^0 \rightarrow  \chi_{1,2}^0 Z$ &24 \%   & &
&&&&&&
\\   
  &$\chi_{3}^0 \rightarrow  \chi_{1,2}^0 h$ &22 \%   & &
&&&&&&
\\   
 \hline \hline
 
    \end{tabular}
   \caption{Dominant production and decay channels for the NLSPs. The mass parameter for NLSPs is taken to be 500 GeV  and $\tan\beta=10$, $\mu>0$, and LSP mass parameters of 100 GeV  is used a benchmark point.  Numbers in parentheses show the variation of the decay branching fractions for $\tan\beta$ varying between 3 to 50.  For signals listed in the last 7 columns, there are always MET + possible soft jets/leptons.   
  }
\label{table:signal} 
\end{table}

\section{Charginos and Neutralinos at the LHC}
\label{sec:pro_decay}

We have laid out the most general chargino and neutralino scenarios based on the relations among the gaugino soft mass parameters $M_{1},\ M_{2}$ and the Higgsino mass parameter $\mu$. In the absence of substantial mixing when all the mass parameters are of the similar size, the three sets of multiplets (namely a Bino, three Winos and four Higgsinos) are each nearly degenerate in mass, respectively. 

The decay patterns for the next to the LSPs (NLSPs)  in the six cases are given in Fig.~\ref{fig:CaseABC_decay}.  The production and decay channels for the three scenarios with six distinctive cases are summarized in Table \ref{table:signal}  \cite{THSPSS}.
For each case, we show the dominant pair production channels for the NLSPs (neutralinos and charginos), and decay modes with branching fractions for various NLSPs as discussed above.  
To guide the searches at the LHC, we combine with the decay branching fractions of the corresponding NLSPs, for each production mode, and show the total branching fraction into each particular final state 
\bea
XY=W^+W^-,\ W^\pm W^\pm,\ WZ,\ Wh,\ Zh,\ ZZ,\ {\rm and}\ hh,
\label{eq:XY}
\eea  
as in Table \ref{table:signal}. 
Note that all of the final states in addition include missing transverse energy introduced by $\chi_1^0$ LSP, as well as soft jets and leptons that might appear from decays between nearly degenerate particles in LSP multiplet.


\begin{figure}
 \minigraph{8.1cm}{-0.2in}{(a)}{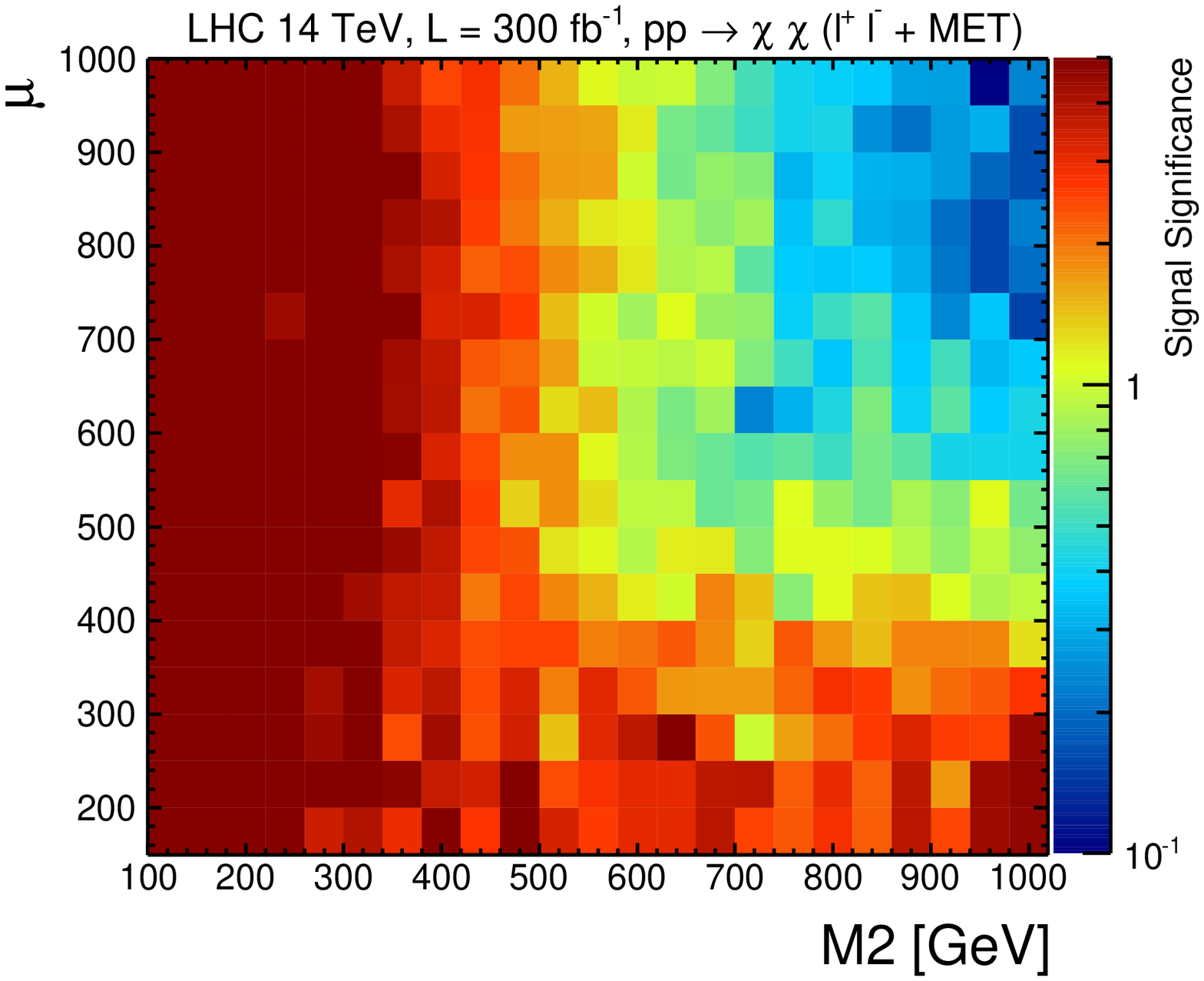}
 \minigraph{8.1cm}{-0.2in}{(b)}{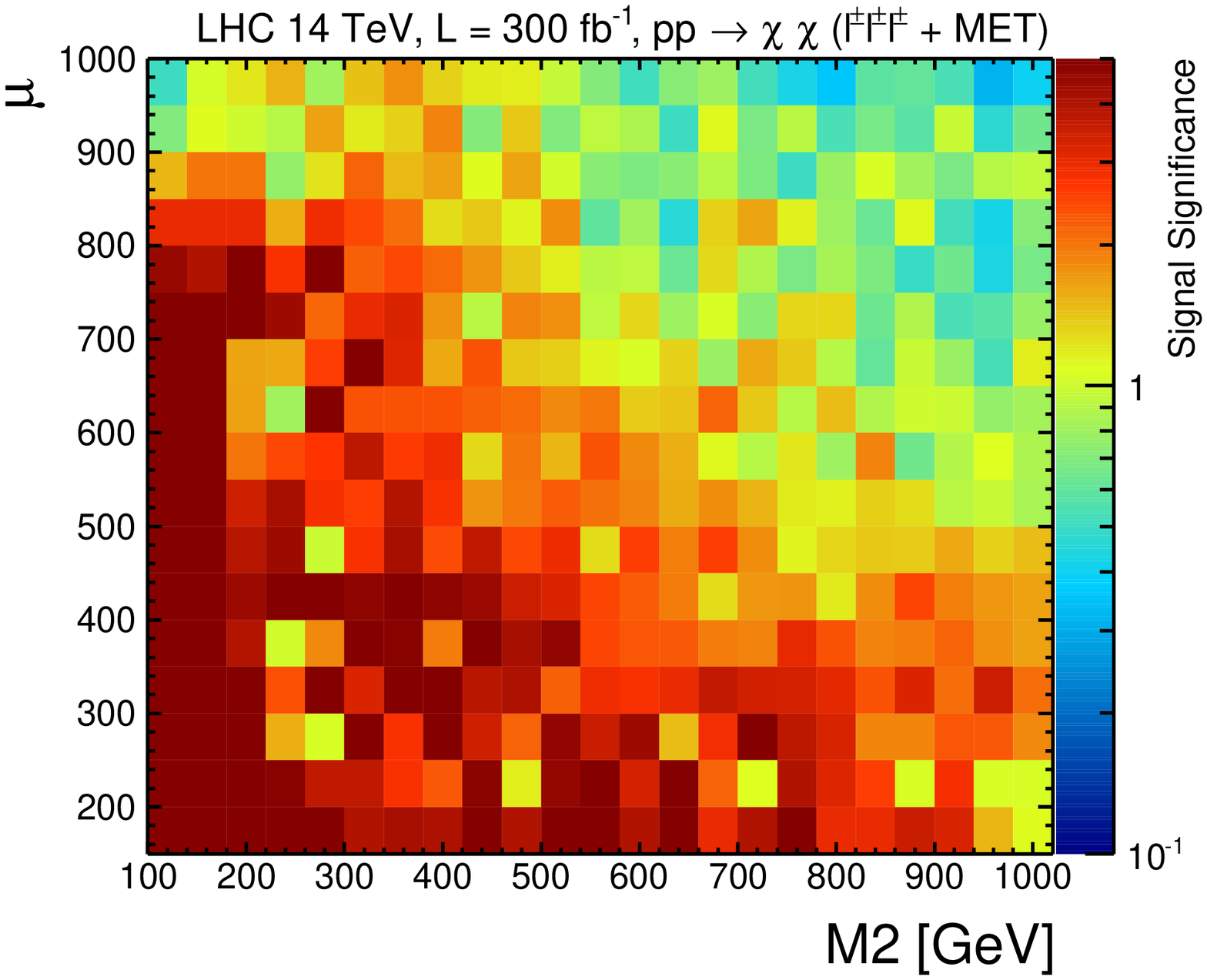}
 \minigraph{8.1cm}{-0.2in}{(c)}{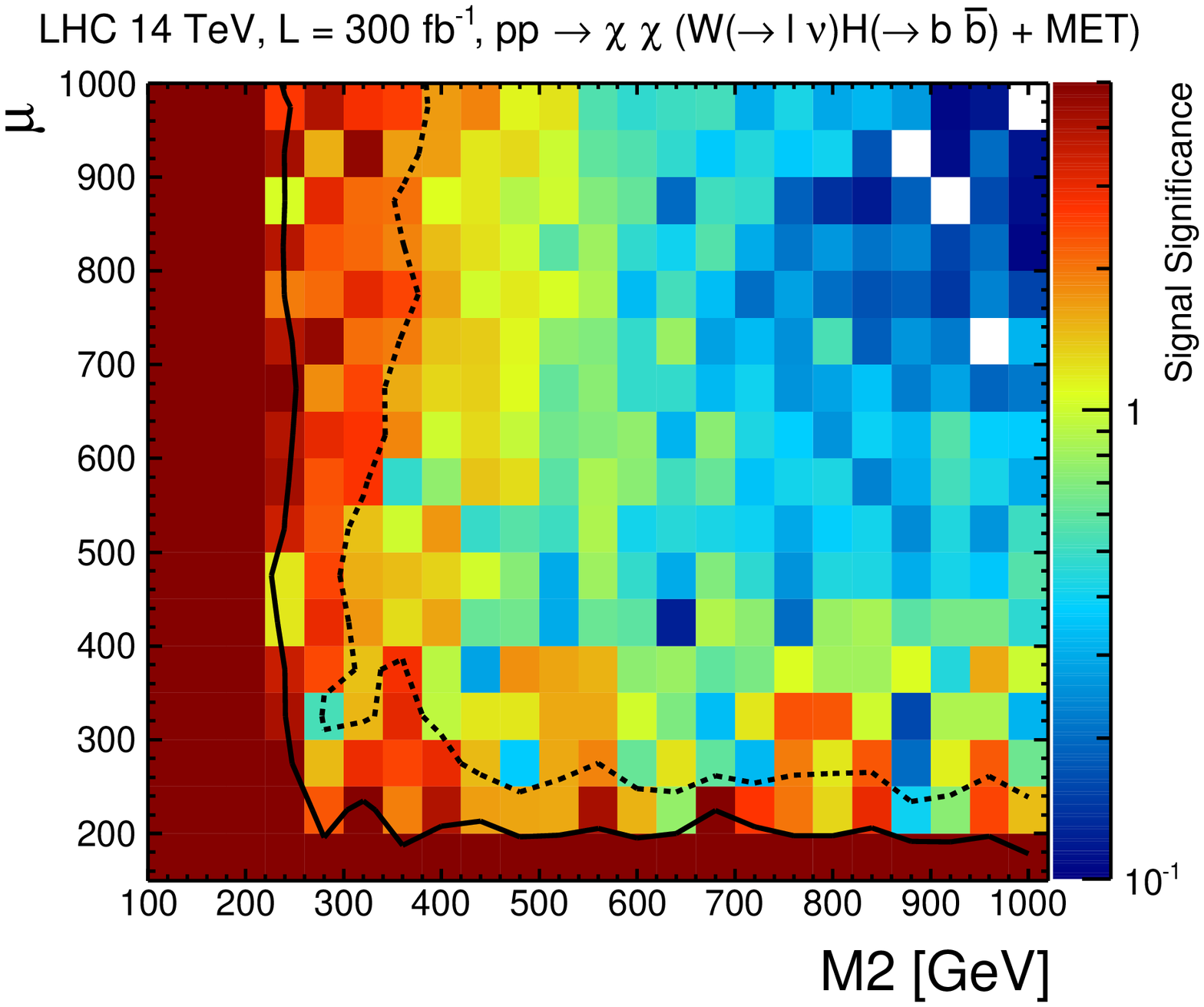}
 \minigraph{8.1cm}{-0.2in}{(d)}{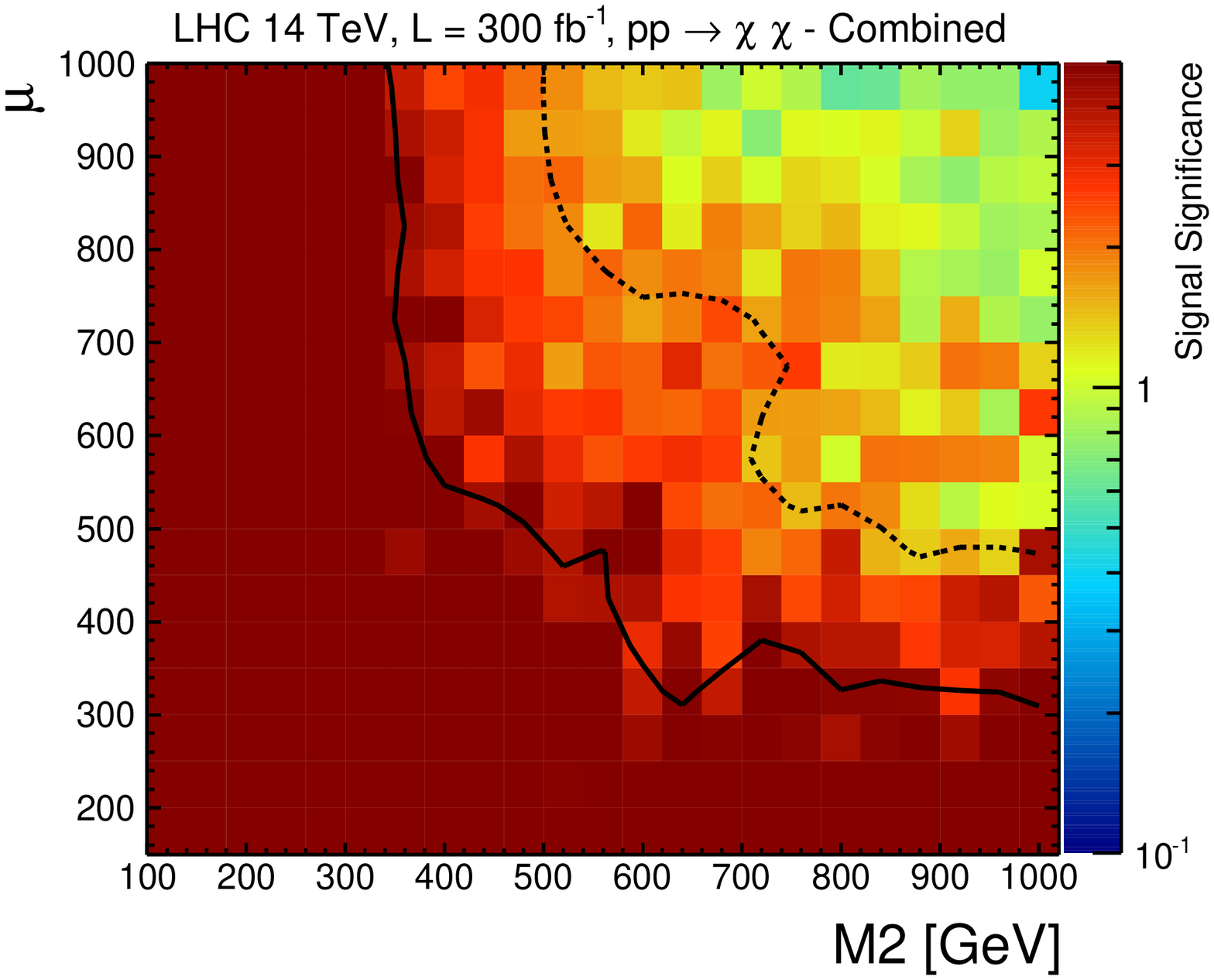}
\caption{
Sensitivity reach at the 14 TeV LHC with 300 fb$^{-1}$  for Case A Bino-like LSP in $\mu-M_{2}$ plane,  
(a) opposite-sign dilepton ($\ell^+\ell^-+\etmiss$), (b) tri-lepton ($\ell\ell\ell+{\rm jets}+\etmiss$),  (c) for $Wh$ ($\ell bb+\etmiss$) and (d) combined sensitivity for all six channels.   The statistical significance is labeled by the color code on the right-hand side. The solid and dashed curves indicate the $5 \sigma$ discovery and  95\% C.L. exclusion   reach. The other MSSM parameters are set to be $M_1=0$ GeV, $\tan\beta=10$ and $\mu>0$.   
} 
\label{fig:LHCreach}
\end{figure}

%
Monte Carlo simulations are used to estimate the SM backgrounds, as well as to calculate the efficiency for various electroweakino 
productions. In this study, events are generated using {\sc MADGRAPH} event generator \cite{madgraph} and {\sc PYTHIA}~\cite{Sjostrand:2006za} for parton shower 
and hadronization. Next-to-leading-order (NLO) cross sections are used for background and signal normalization, calculated using {\sc MCFM}~\cite{Campbell:2010ff} 
and {\sc PROSPINO}~\cite{Beenakker:1996ed}, respectively. For both background and signal samples~\cite{Avetisyan:2013onh}, the events are processed through the
Snowmass detector~\cite{Anderson:2013kxz} using Delphes \cite{deFavereau:2013fsa} parametrized simulation and object reconstruction. Large statistics 
of background samples are generated using the Open Science Grid infrastructure~\cite{Avetisyan:2013dta}. Effects due to additional 
interactions (pile-ups) are studied and they are found to small for 300 fb$^{-1}$ luminosity scenario~\cite{Anderson:2013kxz}. Jets are reconstructed 
using the anti-$k_T$ clustering algorithm~\cite{Cacciari:2008gp} with a distance parameter of 0.5, as implemented in the {\sc FASTJET} 
package~\cite{Cacciari:2011ma}. We have also assumed a systematic uncertainty of $20\%$ in this study.

We present the sensitivity reach at the 14 TeV LHC with 300 fb$^{-1}$ in Fig.~\ref{fig:LHCreach} for  Case A Bino-like LSP in $\mu-M_{2}$ plane.  
We show three representative channels 
(a) opposite-sign dilepton ($\ell^+\ell^-+\etmiss$), (b) tri-lepton ($\ell\ell\ell+{\rm jets}+\etmiss$),  and (c)   $Wh$ ($\ell bb+\etmiss$).  We also show in (d) the combined sensitivity for all six search channels \cite{THSPSS}.   The statistical significance is labeled by the color code on the right-hand side. The solid and dashed curves indicate the $5 \sigma$ discovery and  95\% C.L. exclusion   reach. 
 As expected, we see that the di-lepton mode in (a) is more sensitive to Case AI with $M_{2}< \mu$ with certain sensitivity to low $\mu$ as well, the trilepton mode in (b) is more sensitive to Case AII with $\mu < M_{2}$, while the single lepton plus $h \to b\bar b$ in (c) is mainly sensitive to low $M_{2}$. 
 It is important to note the complementarity of the different channels.
The $Wh\ (h\to b\bar b)$ final state may yield a sensitivity of $95\%$ C.L.~exclusion (5$\sigma$ discovery) to the mass scale $M_{2},\ \mu \sim 250-400$ GeV ($200-250$ GeV).
Combining with all the other decay channels, the $95\%$ C.L.~exclusion (5$\sigma$ discovery) may be extended to $M_{2},\ \mu \sim 480-700$ GeV ($320-500$ GeV).%
 

\begin{figure}
\centering
 \minigraph{8.1cm}{-0.2in}{(a)}{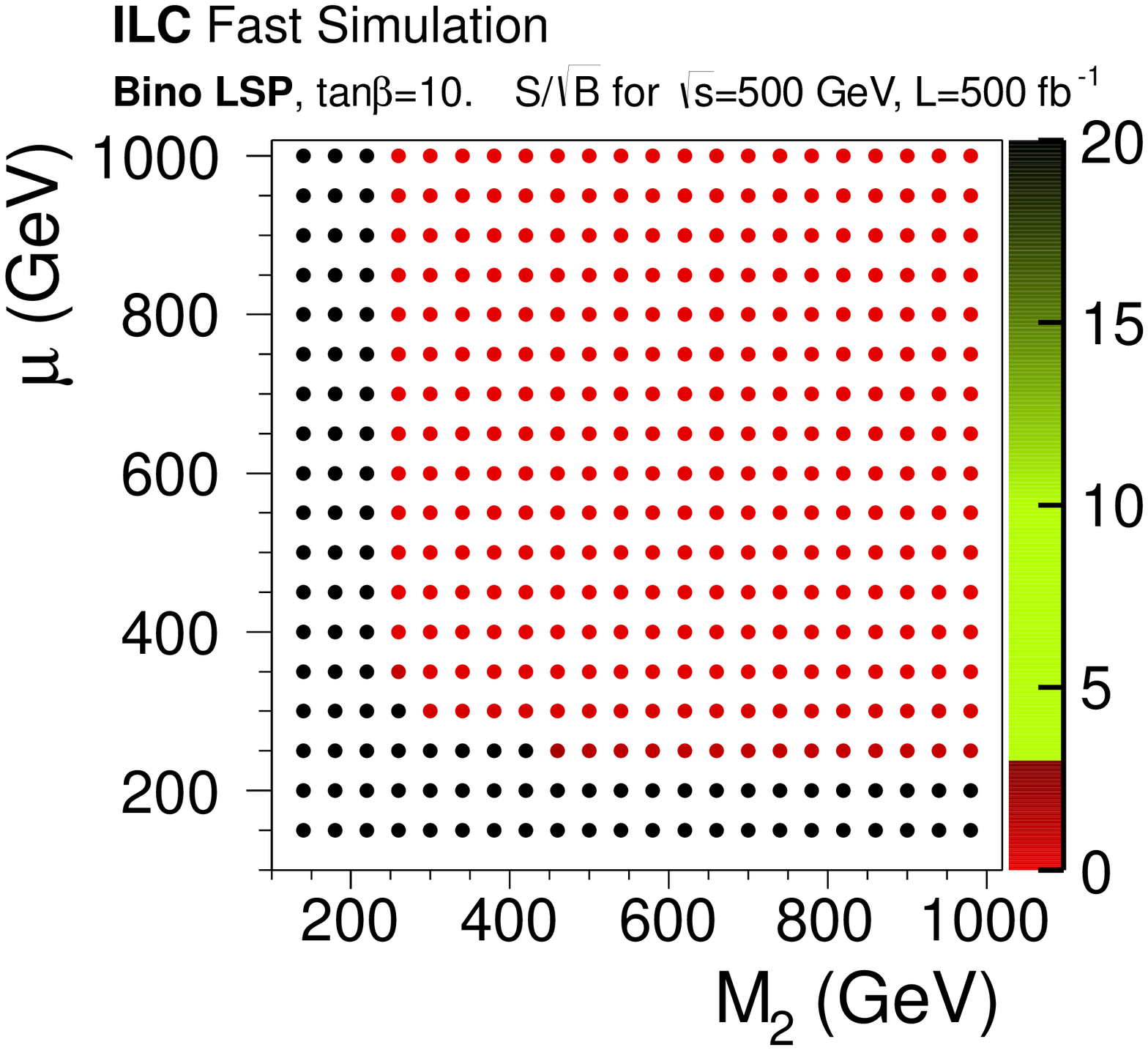}
 \minigraph{8.1cm}{-0.2in}{(b)}{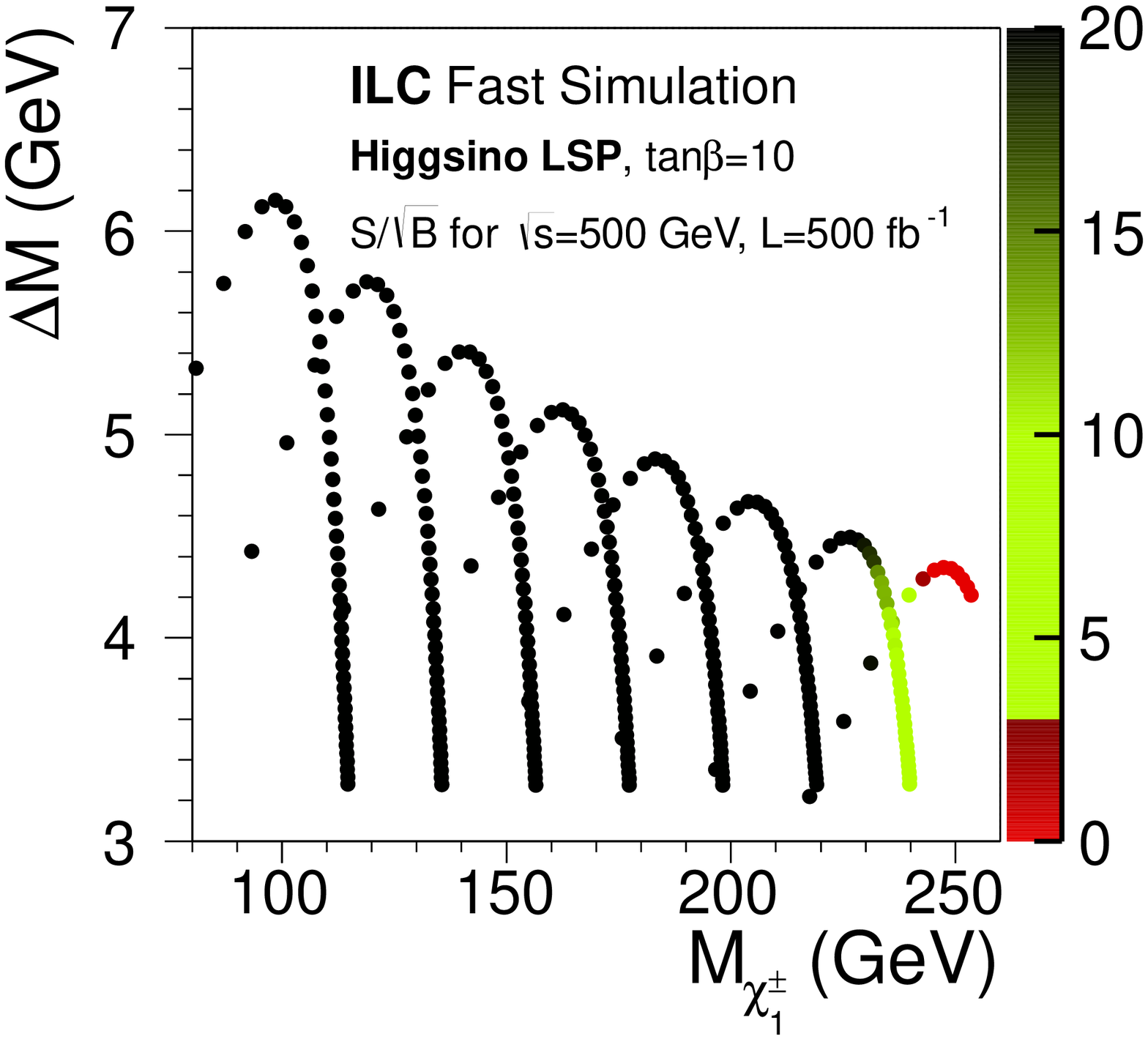}
\caption{Sensitivity reach for electroweakinos at the 500 ILC with 500 fb$^{-1}$ luminosity,  studied with fast simulation: 
  (a) for the Bino-like LSP in $\mu-M_2$ plane, (b) for the case of Higgsino-like LSP in $\Delta M = M_{\chi_1^\pm} -M_{\chi_1^0}$ versus Higgsino mass $M_{\chi_1^\pm}$.  
  The green/black-colored points indicate points with $S/\sqrt{B}>3\sigma$, while the red-colored points are with $S/\sqrt{B}<3 \sigma$.   The statistical significance is labeled by the color code on the right-hand side. 
}
\label{fig:ILCreach}
\end{figure}


\section{Charginos and Neutralinos at the ILC}
\label{sec:ilc}


Due to the rather small electroweak production cross sections and large SM backgrounds at the LHC, the discovery of the charginos and neutralinos via direct 
production would be very challenging as discussed in the previous section. Even if the signal is observed, the determination of their properties would be very difficult. 
On the other hand an ILC will provide major advantages in identifying and determining the underlying properties.
The typical cross sections are of the order of 100 fb. Once crossing the kinematical threshold, the fermionic pair production reaches the maximum rather soon.
In Fig.~\ref{fig:ILCreach}(a), 
we show the signal sensitivity at the 500 GeV ILC with 500 fb$^{-1}$ for the case of Bino-like LSP. We see that the sensitivity reach is about half of the 
center-of-mass energy in the NLSP's mass, be it Wino-like (dots along vertical axis) or Higgsino-like (dots along the horizontal axis). Thus 
for an upgraded ILC with $\sqrt{s}=1\,$TeV, the reach in NLSP mass will extend to $\sim 500\,$GeV.

On the other hand, for the cases of the LSP being Higgsino-like or Wino-like, the multiplet of LSPs could be copiously produced if 
$\mu,\ M_{2} \lesssim \sqrt{s}/2$.   The mass difference between the degenerate LSPs are typically small, in the range of few hundred MeV to a few GeV, resulting in relatively soft decay products in the final states.   The initial state radiation (ISR) can be used to enhance the signal sensitivity.    The expected statistical significance from fast simulation for a Higgsino-like LSP is    shown in Fig.~\ref{fig:ILCreach}(b) in the $\Delta M - M_{\chi_1^{\pm}}$ plane, with $\Delta M = M_{\chi_1^\pm} -M_{\chi_1^0}$, thus 
probing the compressed regions. In addition, if the SUSY signal is indeed observed at the ILC,
it will be possible to separate the signal into the chargino and neutralino components, with mass resolutions and cross section measurements
being typically at the $O(1\%)$ level~\cite{Suehara:2009bj,Berggren:2013vfa}.
In particular for the Higgsino-like LSP case, it has been shown that these measurements yield sufficient precision to put interesting constraints on $M_1$ and $M_2$, 
even if they are in the multi-TeV regime~\cite{Berggren:2013vfa}.

\section{Outlook}
Looking forward, the high luminosity LHC with 3000 fb$^{-1}$ would be expected to extend the 5$\sigma$ electroweakino reach to a mass generically of 800 GeV assuming a $100\%$ branching fraction to the gauge bosons \cite{ATLAS_CMS_3L_future}. 
It would be a pressing issue to address to what extent one would be able to uncover the observationally difficult scenarios where the lower lying 
electroweakinos are in a compressed LSP spectrum and the NLSPs may not be copiously produced at the LHC. 
Furthermore, if a multiple TeV lepton collider is ever available \cite{Lebrun:2012hj,Alexahin:2013ojp}, it would readily cover to a mass scale about a half of the center of mass energy. 

 \acknowledgments
We thankfully acknowledge the support by the DFG through the SFB 676 ``Particles, Strings and the Early Universe''. The work of T.H.~was supported in part by  the U.S.~Department of Energy  under grant DE-FG02-95ER40896 and in part by the PITT PACC, the work of S.P.~was supported in part by the Department of Energy grant DE-FG02-90ER40546 and the FNAL LPC Fellowship, the work of S.S.~was supported by the  Department of Energy  under Grant DE-FG02-04ER-41298, and that of T.T~was in  part supported by the Grant-in-Aid for Specially Promoted Research No.~23000002 by the Japan Society for Promotion of Science.
\raggedright

\end{document}